\documentclass[english,apm, reprint, twocolumn, superscriptaddress,showpacs,amsmath,amssymb]{revtex4-1}
\usepackage{lmodern}
\usepackage[T1]{fontenc}
\usepackage[latin9]{inputenc}
\usepackage{booktabs}
\usepackage{textcomp}
\usepackage{amssymb}
\usepackage{graphicx}

\makeatletter

\providecommand{\tabularnewline}{\\}

\@ifundefined{textcolor}{}
{%
 \definecolor{BLACK}{gray}{0}
 \definecolor{WHITE}{gray}{1}
 \definecolor{RED}{rgb}{1,0,0}
 \definecolor{GREEN}{rgb}{0,1,0}
 \definecolor{BLUE}{rgb}{0,0,1}
 \definecolor{CYAN}{cmyk}{1,0,0,0}
 \definecolor{MAGENTA}{cmyk}{0,1,0,0}
 \definecolor{YELLOW}{cmyk}{0,0,1,0}
}

\usepackage{upgreek}

\newcommand{\SSOO}{Sr$_2$ScOsO$_6$}

\newcommand {\TC} {$T_{\mathrm{C}}$}

\newcommand {\TN} {$T_{\mathrm{N}}$}

\makeatother

\usepackage{babel}
\begin{document}

%

\title{Magnetic order and electronic structure of $5d^{3}$ double perovskite \SSOO{}}

\author{A. E. Taylor }

\affiliation{Quantum Condensed Matter Division, Oak Ridge National Laboratory,
Oak Ridge, TN 37831, USA}

\author{R. Morrow}

\affiliation{Department of Chemistry, The Ohio State University, Columbus, OH
43210-1185, USA}

\author{D. J. Singh}

\affiliation{Materials Science and Technology Division, Oak Ridge National Laboratory, Oak Ridge, TN 37831, USA}

\author{S. Calder}

\affiliation{Quantum Condensed Matter Division, Oak Ridge National Laboratory,
Oak Ridge, TN 37831, USA}

\author{M. D. Lumsden}

\affiliation{Quantum Condensed Matter Division, Oak Ridge National Laboratory,
Oak Ridge, TN 37831, USA}

\author{P. M. Woodward}

\affiliation{Department of Chemistry, The Ohio State University, Columbus, OH
43210-1185, USA}

\author{A. D. Christianson}

\affiliation{Quantum Condensed Matter Division, Oak Ridge National Laboratory,
Oak Ridge, TN 37831, USA}

\affiliation{Department of Physics and Astronomy, University of Tennessee, Knoxville,
TN 37996, USA}


\begin{abstract}
The magnetic susceptibility, crystal and magnetic structures, and electronic structure of double perovskite \SSOO{} are reported. Using
both neutron and x-ray powder diffraction we find that the crystal
structure is monoclinic \emph{P}2$_{1}/$\emph{n}
from 3.5 to 300\,K. Magnetization measurements
indicate an antiferromagnetic transition at $T_{\mathrm{N}}=92\,$K,
one of the highest transition temperatures of any double perovskite hosting
only one magnetic ion. Type I antiferromagnetic order is determined by neutron powder diffraction, with an Os moment
of only 1.6(1)$\,\upmu_{\mathrm{B}}$, close to half the spin-only
value for a crystal field split 5$d$ electron state with $t_{2g}$$^{3}$
ground state. Density functional calculations show that this reduction is largely the result of strong Os-O hybridization, with spin-orbit coupling responsible for only a $\sim0.1\,\upmu_\mathrm{B}$ reduction in the moment. 
\end{abstract}

\maketitle


It is important to understand the nature of the $d^{3}$ electronic
state, as it appears to produce the highest magnetic transition
temperatures found in perovskite systems~\cite{middey_route_2012}.
Perovskites NaOsO$_{3}$, $5d^{3}$, and SrTcO$_{3}$, $4d^{3}$,
have $T_{\mathrm{N}}=411\,$K and $T_{\mathrm{N}}\approx1000\,$K,
respectively~\cite{shi_continuous_2009,calder_magnetically_2012,rodriguez_high_2011,thorogood_structural_2011}.
In the double perovskite Sr$_{2}$CrOsO$_{6}$ two $d^{3}$ ions are
present, Cr$^{3+}$ $3d^{3}$ and Os$^{5+}$ $5d^{3}$, and a ferrimagnetic
transition with $T_{\mathrm{C}}=725\,$K is found, the highest known
transition temperature in any double perovskite~\cite{krockenberger_sr2croso6:_2007}.
However, the presence of two magnetic ions in Sr$_{2}$CrOsO$_{6}$
makes the interactions controlling this high-\TC{} insulator more
difficult to unravel~\cite{meetei_theory_2013}. It is clear that
the influence of the Os $5d^{3}$ electrons must be understood in
order to ultimately understand the high-\TC{} found in Sr$_{2}$CrOsO$_{6}$. 

The $d^{3}$ electronic configuration in
a perovskite-type structure is, at first sight, an unlikely host of
complex magnetic behavior. The octahedral environment splits the $d$
orbitals into $t_{2g}$ and $e_{g}$ levels, stabilizing a half occupied
$t_{2g}$$^{3}$ configuration. This is normally assumed to be fully
orbitally quenched, resulting in a relatively classical $S=3/2$ moment
behavior~\cite{chen_spin-orbit_2011}. However, many investigations
in $4d^{3}$ and $5d^{3}$ compounds have found that the ordered
magnetic moment measured by neutron diffraction is significantly reduced
from the spin only value of $\mu=3\,\upmu_{\mathrm{B}}$~\cite{aczel_frustration_2013,aczel_coupled_2013,battle_crystal_1989,calder_magnetic_2012,calder_magnetically_2012,paul_magnetically_2015,kayser_crystal_2014}.
 Explanations of this reduced moment have centered on covalency and frustration. It has recently been suggested, however,
that the effect of spin-orbit coupling (SOC) cannot be ignored in $4d^{3}$ and $5d^{3}$ systems~\cite{matsuura_effect_2013}.
In Ref.~\cite{matsuura_effect_2013} they find that the electronic state may lie between
L-S and J-J coupling regimes, and therefore the $t_{2g}$$^{3}$ assumption
is invalid. Recent experimental investigations have therefore discussed the role of SOC in $4d^3$ and $5d^3$ double perovskites~\cite{aczel_frustration_2013, aczel_exotic_2014, carlo_spin_2013, kermarrec_frustrated_2014, nilsen_diffuse_2015}. 

Here we investigate the magnetic insulator \SSOO{}, which has one of the
highest known magnetic transition temperatures in double perovskites
hosting only one magnetic ion~\cite{vasala_a2bbo6_????,yuan_high-pressure_2015}. It is the $3d^{0}-5d^{\ensuremath{3}}$
analogue of $3d^{3}-5d^{3}$ Sr$_{2}$CrOsO$_{6}$, and thus serves as a model system to unravel the behavior of $d^{3}$ Os$^{5+}$. 
Sc ions strongly favor the nonmagnetic 3+ oxidation state.
X-ray absorption
measurements are consistent with an Os$^{5_{+}}$ valence state and
large 3.6\,eV $t_{2g}$ to $e_{g}$ splitting in \SSOO{}~\cite{choy_study_1998}.
We confirm an antiferromagnetic (AFM) like peak in the susceptibility~\cite{paul_magnetically_2015},
and by neutron powder diffraction (NPD) we show that this is the result
of type I AFM ordering at $T_{\mathrm{N}}=92\,$K.
The magnetic moment on the osmium site is $1.6\pm0.1\,\upmu_{\mathrm{B}}$.
This is significantly reduced from the spin-only value, as is the
case for NaOsO$_{3}$ [$\mu_{\mathrm{Os}}=1.0(1)\,\upmu_{\mathrm{B}}$]
and SrTcO$_{3}$ [$\mu_{\mathrm{Tc}}=1.87(4)\,\upmu_{\mathrm{B}}$],
Sr$_{2}$CrOsO$_{6}$ [$\mu_{\mathrm{Os}}=0.7(3)\,\upmu_{\mathrm{B}}$]
and other $4d$ and $5d$ double perovskites. Density functional theory (DFT) calculations show
that the reduction in moment in \SSOO{} is largely the result of strong Os--O hybridization, with SOC responsible for only a small fraction of the reduction. 



Polycrystalline \SSOO{} was synthesized by combining stoichiometric
quantities of SrO$_{2}$, Os, OsO$_{2}$ and Sc$_{2}$O$_{3}$. Ground
mixtures were contained in alumina tubes
and sealed in evacuated silica vessels for heatings of 48\,hours
at 1000\,\textdegree C. This was followed by regrinding and identical
reheating an additional three times. 

Magnetization measurements were conducted using a \textsc{mpms squid}
magnetometer, from 5 to 400\,K under an applied field of 1\,kOe. Laboratory x-ray powder diffraction
(XRPD) measurements were conducted at room temperature on a Bruker D8 advance using a Ge(111) monochromator and a Cu radiation
source ($\lambda=1.54056\,\mathrm{\AA}$).       

NPD measurements were conducted on \textsc{powgen} at the Spallation
Neutron Source at Oak Ridge National Laboratory (ORNL)~\cite{huq_powgen_2011}.
The 0.906\,g sample was contained in a vanadium can and measured
at 10\,K and 300\,K.
The chopper settings were chosen to correspond to a measured $d$-spacing
range of 0.2760--3.0906\,$\mathrm{\AA}$. NPD and XRPD data were analyzed
using the Rietveld method as implemented in \textsc{gsas+expgui}~\cite{larson_general_1994,toby_expgui_2001}. 
No impurity peaks were identified in the diffraction patterns for this sample. 

To determine the magnetic structure, separate NPD measurements were
performed using a 6\,g sample on HB-2A
at the High Flux Isotope Reactor at ORNL~\cite{garlea_high-resolution_2010}.
The sample was sealed in a vanadium can with He exchange gas, and
a closed cycle refrigerator was used to reach 3.5\,K.
Measurements were conducted with a neutron wavelength of 2.4137$\,\mathrm{\AA}$,
with collimation of 12$^\prime$--open--12$^\prime$. The resulting
data were analyzed with the Rietveld refinement suite Fullprof \cite{rodriguez-carvajal_recent_1993},
using the magnetic form factor for Os$^{5+}$ from Ref.~\cite{kobayashi_radial_2011}. An unidentified trace impurity is present in this sample, which is too small to affect the results of the magnetic refinement.

\begin{figure}
\includegraphics[clip,width=0.95\columnwidth]{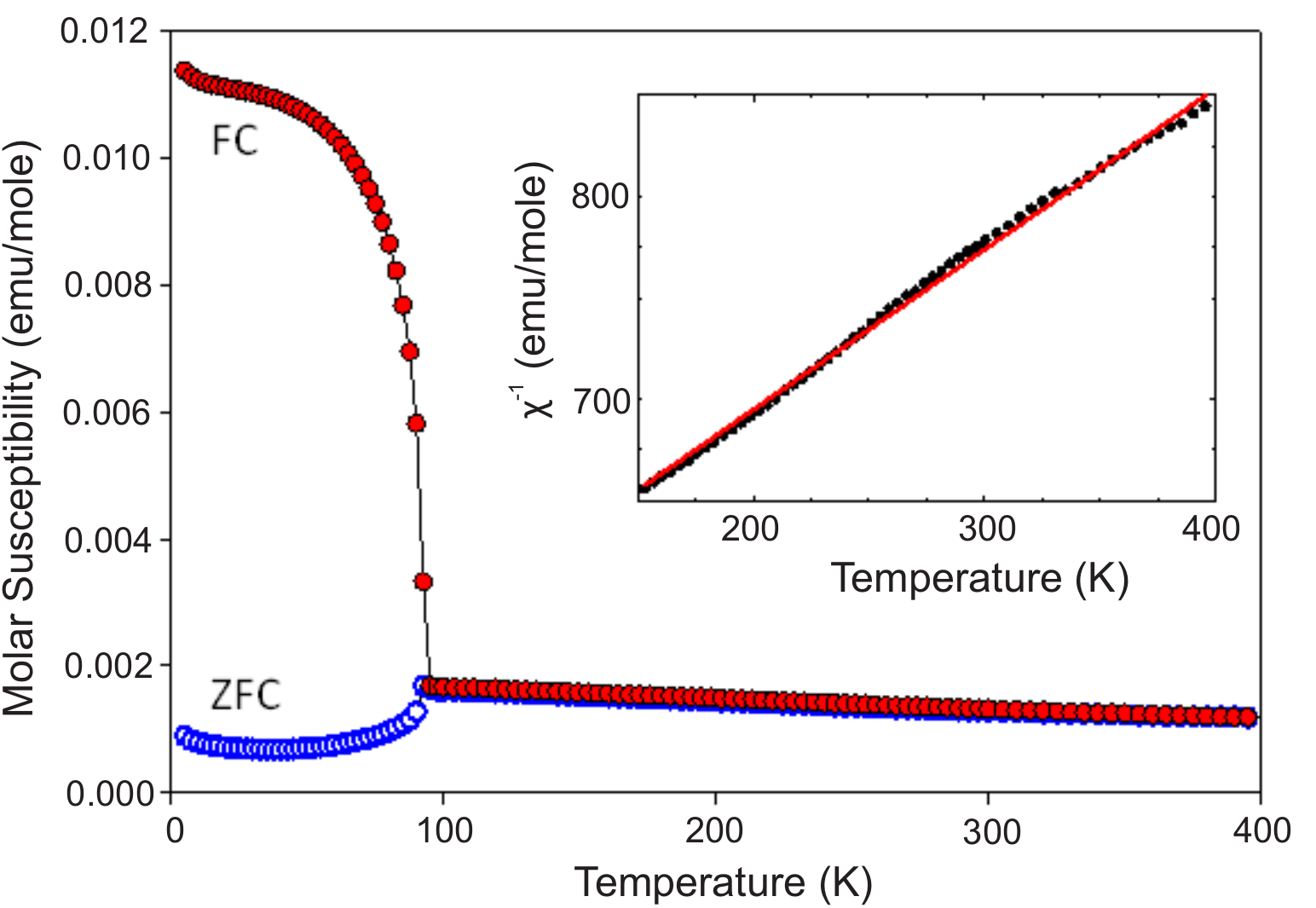}

\protect\caption{\label{fig:Magnetization}(Color online) Temperature dependence of
the susceptibility measured in a 1\,kOe applied field.
Blue open symbols and red closed symbols show the zero field cooled (ZFC)
and field cooled (FC) [at 1\,kOe] measurements, respectively. The inset shows the inverse
susceptibility well above \TN{}, and the result of a Curie-Weiss fit to the data (solid line). }
\end{figure}

A peak at 92.5$\,$K is seen in the ZFC susceptibility
data, see Fig.~\ref{fig:Magnetization}, which is characteristic of
AFM order. The FC data shows a ferromagnetic component,
that is likely the result of a small canting of the magnetic moments.
At 5\,K the ZFC M versus H measured up to 5\,T (not shown) shows a linear
relation consistent with AFM order. These results are similar to the
magnetization measurements recently reported in Ref.~\cite{paul_magnetically_2015}.
Earlier magnetization measurements show a peak in the susceptibility at ~25\,K, which may
be due to different sample quality, as a cubic symmetry was determined from XRPD~\cite{choy_study_1998}. 
The inverse susceptibility versus temperature
in the paramagnetic region for our sample is shown in the inset to
Fig.~\ref{fig:Magnetization}, along with a Curie-Weiss fit to the
data. The fit yields a Curie-Weiss temperature $\Theta=-677\,$K and
an effective moment of $\mu_{eff}=3.18\,\upmu_{\mathrm{B}}$. The
reduction in the effective moment from the expected spin-only value
of $\mu_{eff}=3.87\,\upmu_{\mathrm{B}}$ (assuming a g-factor of 2) is likely an indication of
SOC acting as a perturbation on the $t_{2g}$$^{3}$
state. The frustration index for \SSOO{}
is $|\Theta|/T_{\mathrm{N}}=7.4$, indicating a high degree of frustration
as expected for a near face-centered-cubic arrangement of magnetic
ions. 

\begin{figure}
\includegraphics[clip,width=0.93\columnwidth]{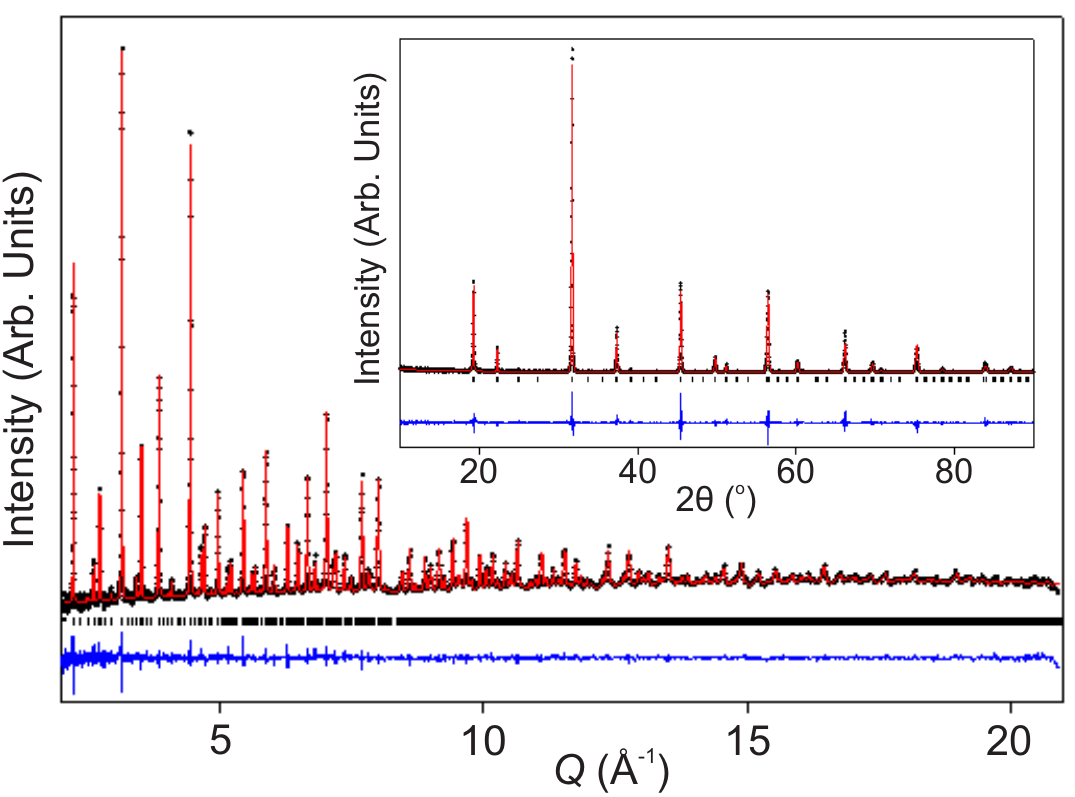}
\protect\caption{\label{fig:Diffraction_Struct} (Color online) Rietveld refinements
against data (black crosses), collected at room
temperature by NPD on \textsc{powgen} (main panel) and XRPD (inset). The difference between calculation and data is also shown,
and is offset for clarity. }
\end{figure}

The results of XRPD measurements and Rietveld analysis are presented
in the inset to Fig.~\ref{fig:Diffraction_Struct}. As Sc and Os have significantly
different atomic masses, x-rays are particularly sensitive to anti-site
mixing between $B$ and $B^\prime$ sites. We find that the sites
are fully occupied, with 0.952(4) Os {[}Sc{]} and 0.048(4) Sc {[}Os{]}
occupancy on the $B^\prime$ {[}$B${]} site. Some anti-site disorder is unsurprising given that the ratio
of  ionic radii of Sc$^{3+}$/Os$^{5+}$  is smaller than 100\,\% ordered
Os double perovskites \cite{paul_magnetically_2015,morrow_independent_2013}.
The occupancies determined from XRPD were subsequently used in the
NPD structural refinements. NPD is advantageous for a full structural
refinement due to a greater sensitivity to the oxygen positions
in the system. 

The data from NPD on \textsc{powgen} at 300$\,$K are shown in the main panel of Fig.~\ref{fig:Diffraction_Struct}.
While the metric of the unit cell was highly pseudocubic, the space
group symmetry was determined by the presence of X point reflections
in the NPD pattern (two odd indices and one even as indexed on a 2$\times$2$\times$2
perovskite cell)~\cite{barnes_structure_2006} as the common monoclinic
$P2_{1}/n$~\cite{lufaso_structure_2006} with the tilt system $a^{-}a^{-}c^{+}$~\cite{glazer_classification_1972}.
The Wykcoff positions of the atoms are: Sc on 2$d$ (\textonehalf{}
0 0), Os on 2$c$ (0 \textonehalf{} 0), and Sr, O1, O2 and O3 on 4$e$
sites ($x$ $y$ $z$). Due to the high degree of pseudosymmetry,
there was reduced sensitivity to the refinement of the oxygen positional
parameters. Soft constraints were therefore placed on the bond lengths
with the weakest weighting, using values of 2.06 and 1.96$\,\mathrm{\AA}$
in accordance with Sc--O and Os--O bond lengths found in similar
double perovskites~\cite{barnes_structure_2006,krockenberger_sr2croso6:_2007}.
The introduction of soft constraints results in only a marginal difference in the quality of the refinement (R$_\mathrm{wp}$ is increased by only 0.02 to 4.32\,\%).  However, if the soft constraints are removed unphysical bond lengths result. We find that space group $P2_{1}/n$ describes
the structure down to 3.5\,K, with no evidence for any structural phase transition. Previous reports of the crystal structure were only available
from XRPD at room temperature \cite{choy_study_1998,paul_magnetically_2015}.
Our results agree with the space group $P2_{1}/n$ reported in Ref.~\cite{paul_magnetically_2015},
but we find the monoclinic angle $\beta=90.218(3)^{\circ}$ to be
significantly larger than their result, $\beta=90.083(2)^{\circ}$.
The complete structural results from the Rietveld refinements for
300\,K and 10\,K are given in Table~\ref{tab:Diffraction} and~\cite{supplementary}. 

\begin{table}
\resizebox{0.63\columnwidth}{!}{
\begin{tabular}{ccc}
\toprule 
Temperature (K) & 10\,K & 300\,K\tabularnewline
\midrule 
Space Group & \emph{P}2$_{1}/$\emph{n} & \emph{P}2$_{1}/$\emph{n}\tabularnewline
$a\,(\mathrm{\AA})$ & 5.6398(2) & 5.6465(2)\tabularnewline
$b\,(\mathrm{\AA})$ & 5.6373(2)  & 5.6477(2) \tabularnewline
$c\,(\mathrm{\AA})$ & 7.9884(3) & 8.0037(3)\tabularnewline
$V\,(\mathrm{\AA}^{3})$ & 253.98(2) & 255.24(2)\tabularnewline
$\beta$ & 90.219(2)$^{\circ}$ & 90.218(3)$^{\circ}$\tabularnewline
R$_{\mathrm{wp}}$ & 5.41\,\% & 4.32\,\%\tabularnewline
Sr $x$ & 0.004(1)  & 0.004(2)\tabularnewline
Sr $y$ & -0.0142(5)  &  -0.006(2)\tabularnewline
Sr $z$ & 0.2506(8)  &  0.250(1)\tabularnewline
O1 $x$ & 0.2202(7)  & 0.232(1)\tabularnewline
O1 $y$ & 0.2295(7)  & 0.220(1)\tabularnewline
O1 $z$ & 0.0270(7)  & 0.0257(8)\tabularnewline
O2 $x$ & 0.245(1)  & 0.2409(9)\tabularnewline
O2 $y$ & 0.233(1) & 0.2523(9)\tabularnewline
O2 $z$ & 0.4757(9) & 0.4836(9)\tabularnewline
O3 $x$ & 0.5443(8) & 0.5423(9)\tabularnewline
O3 $y$ & 0.006(1) & -0.007(2)\tabularnewline
O3 $z$ & 0.2558(4) & 0.2454(3)\tabularnewline
\bottomrule
\end{tabular}}
\protect\caption{Results from Rietveld refinements of NPD data from \textsc{powgen}, as described in the text.  \label{tab:Diffraction}}
\end{table}

%

\begin{figure*}
\includegraphics[clip,width=1.5\columnwidth]{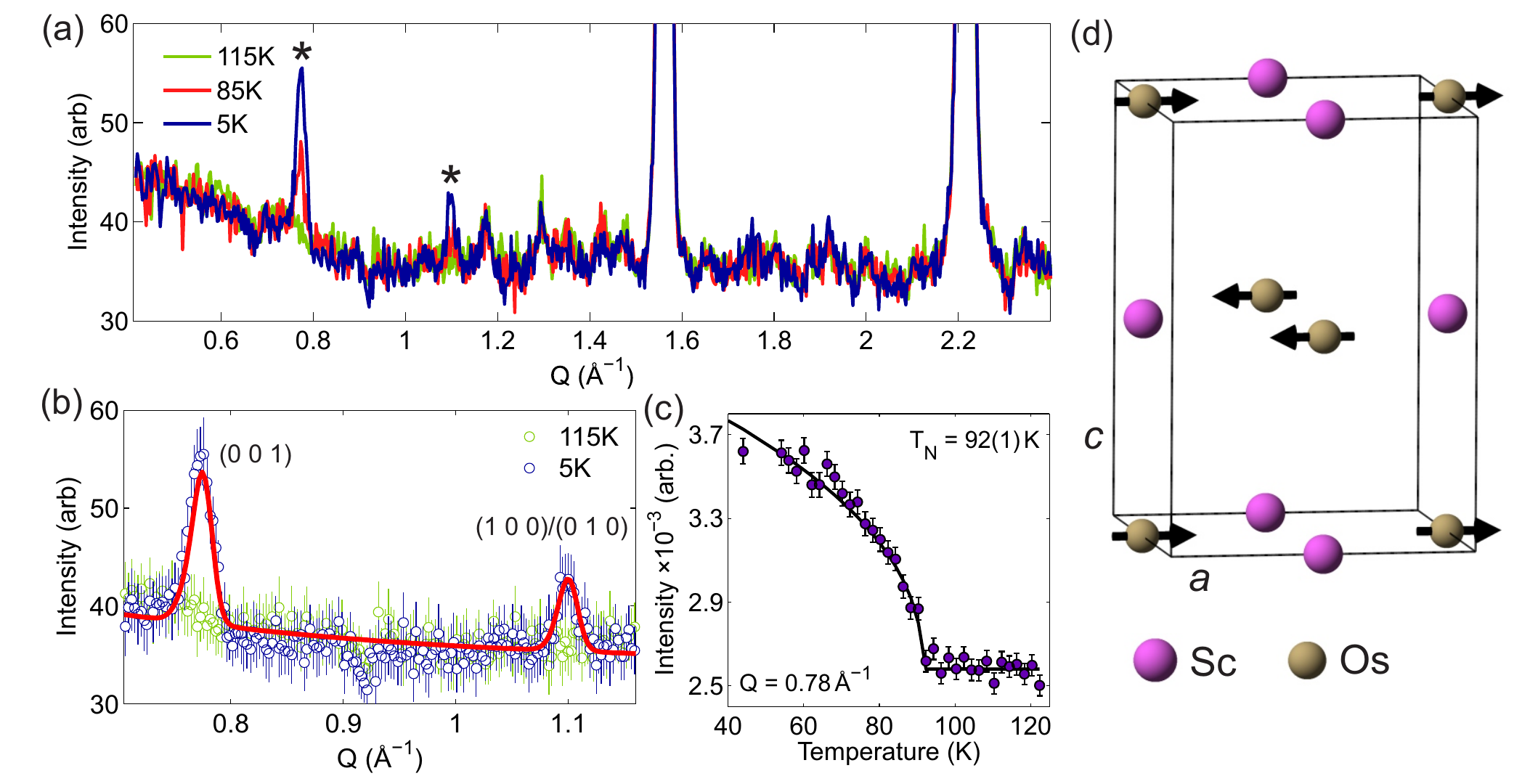}
\protect\caption{\label{fig:Magnetic_diff} (Color online) (a)~NPD data collected with HB-2A at 115\,K ($>$\TN{}),
and 85 and 3.5\,K ($<$\TN{}). Magnetic peaks are indicated by stars.  (b)~Results of Rietveld refinements
of the 3.5\,K data (solid line). (c)~Intensity of scattering at $Q=0.78\,\mathrm{\protect\AA}$
against temperature (circles). The line is the result of a power-law
fit against the data. The intensity is number of counts measured with
the $Q=0.78\,\mathrm{\protect\AA}$ peak counted for 10 minutes per point. (d) The magnetic structure
in one unit cell of \SSOO{}, showing only Os and Sc ions. Moments are depicted along $a$, but their direction within the $a$-$b$ plane is unknown. }
\end{figure*}

Having determined the crystal structure of \SSOO{}, we now turn to
discussion of the magnetic structure. Figure~\ref{fig:Magnetic_diff}(a)
shows data collected on HB-2A at temperatures both above \TN{} (115\,K)
and below \TN{} (85\,K and 3.5\,K). Peaks at $Q=0.78$ and $1.1\,\mathrm{\AA}$
can be seen to develop in the $T<T_\mathrm{N}$ data sets, which are not
present at 115\,K. These reflections can be indexed as (0\,0\,1) and
(1\,0\,0) or $(0\,1\,0)$ respectively, indicating a magnetic propagation
vector $\mathbf{k}=(0\,0\,0)$. There are two irreducible representations in this case, $\Gamma_1$ and $\Gamma_3$,
compatible with the\emph{ P}2$_{1}/$n symmetry with
Os on the 2$c$ Wyckoff site, each representing a simple type I AFM structure. 
Type I order is favored in double perovskites when the nearest neighbor (NN) Os-O-O-Os exchange is dominant,
type II order results if the next nearest neighbor (NNN) Os-Sc-Os
interaction is stronger, as found in some $3d$ ion compounds~\cite{battle_structural_1999,vasala_a2bbo6_????}.

The type I magnetic model with moments in the $a$-$b$ plane was fit to the data,  with the result
shown in Fig.~\ref{fig:Magnetic_diff}(b). As the monoclinic distortion
is small, it is not possible to determine the direction of the magnetic
moment in the $a$-$b$ plane from the powder data. The magnetic structure
is shown in Fig.~\ref{fig:Magnetic_diff}(d) with the spins represented
along the $a$-axis. The size of the Os moment extracted from the
fit is $\mu_{\mathrm{Os}}=1.6(1)\,\upmu_{\mathrm{B}}$, assuming a
95$\,$\% Os occupancy of the $B'$ site is contributing to the scattering.
Significant covalency may explain the observation of a greatly reduced moment determined
by neutron scattering as compared to $\mu_{\mathrm{eff}}$ from magnetization
measurements~\cite{hubbard_covalency_1965}. The detailed temperature
evolution of the $Q_{(001)}=0.78\,\mathrm{\AA}$ peak is shown in Fig.~\ref{fig:Magnetic_diff}(c),
along with a power-law curve that was fit to the data. This confirms
the \TN{}$=92(1)\,$K transition temperature associated with this
magnetic peak. 

It is informative to put these results in context against other double
perovskites which host a $d^{0}d^{3}$ configuration. Despite high
frustration indexes $|\Theta|/T_{\mathrm{N}}$, most of these systems
overcome frustration to order in a type I AFM state at low temperature
{[}see Fig.~\ref{fig:Magnetic_diff}(d){]}, although some show short
range incommensurate order or type III order \cite{battle_crystal_1989,kermarrec_frustrated_2014,aczel_exotic_2014,thompson_long-range_2014,paul_magnetically_2015,battle_crystal_1984}.
\SSOO{}, Sr$_{2}$YOsO$_{6}$ and Ba$_{2}$YOsO$_{6}$ all host long
range type I AFM order, with \TN{}s of 95\,K, 53\,K and 69\,K,
and frustration indexes $|\Theta|/T_{\mathrm{N}}=7.4$, $6.4$ and
$11$ respectively~\cite{kermarrec_frustrated_2014,paul_magnetically_2015}.
Ba$_{2}$YOsO$_{6}$ is the only one of these compounds to maintain
a cubic symmetry, which likely explains the higher level of frustration
found in this system~\cite{kermarrec_frustrated_2014}. \SSOO{}
and Ba$_{2}$YOsO$_{6}$ have the same size magnetic moment as
determined by neutron diffraction, 1.6(1)$\,\upmu_{\mathrm{B}}$ and
1.65(6)$\,\upmu_{\mathrm{B}}$, and so it seems likely that the Os
$d$ electrons are in the same state in each of these materials, explaining
the relatively high \TN{}s observed in both. It would seem that the
monoclinic distortion in \SSOO{} relieves frustration and, along
with a smaller unit cell parameter, allows this material to host the
highest ordering temperature. The presence of a small amount of Sc/Os site disorder in \SSOO{} may also act to enhance \TN{}, as the interactions between  anti-site Os ions and regular site Os neighbors will be strongly AFM. 

It is unclear, however, why Sr$_{2}$YOsO$_{6}$ has the lowest \TN{}
of the three compounds. Its lattice is intermediate
in size between \SSOO{} and Ba$_{2}$YOsO$_{6}$, which in principle
implies strengthened NN interactions over Ba$_{2}$YOsO$_{6}$. Also,
its magnetic moment is larger at 1.91(3)$\,\upmu_{\mathrm{B}}$~\cite{paul_magnetically_2015}
which should energetically favor a higher \TN{}. In the monoclinic symmetry the 12 NN Os distances are not equal, and in  Sr$_{2}$YOsO$_{6}$ the shortest distance is between in-plane ferromagnetically aligned pairs, whereas in \SSOO{} the shortest distance is for out-of-plane AFM aligned pairs. This might explain some lowering of \TN{} in  Sr$_{2}$YOsO$_{6}$, but does not explain the larger magnetic moment. This raises the possibility that a single mechanism both causes a reduction in $\mu$ and enhances the exchange interactions and therefore increases \TN{}, as reflected in \SSOO{} and Ba$_{2}$YOsO$_{6}$. The similar Ru-based compounds  Ba$_{2}$YRuO$_{6}$ and Sr$_{2}$YRuO$_{6}$ have larger moments of $\sim2\,\mu_\mathrm{B}$ and lower \TN{}s of $\sim30\,$K~\cite{battle_crystal_1989,aharen_magnetic_2009,battle_crystal_1984}, whereas high-\TN{} NaOsO$_3$ also has a reduced moment~\cite{calder_magnetically_2012}. Increased hybridization between Os 5$d$ orbitals and O 2$p$ orbitals could be responsible, as this reduces the localized Os moment, but could increase the exchange coupling along the nearest neighbor Os-O-O-Os pathway. 

To investigate the effect of hybridization in \SSOO{}, we have performed DFT calculations using the experimental
crystal structure. The calculations were done using the
general potential linearized augmented planewave (LAPW) method,
\cite{singh-book} as implemented in the WIEN2k code~\cite{wien2k}.
We used the generalized gradient approximation (GGA) of Perdew, Burke and Ernzerhof~\cite{pbe}, 
both by itself and including a Coulomb repulsion via the GGA+U scheme, with $U$=3\,eV.
SOC was included in the calculations. We find an AFM ground state in
agreement with the experimental observation, and confirm that covalency is largely responsible for the reduction of the Os moment well below the $t_{2g}$$^3$ spin-only value. 

\begin{figure}
 \includegraphics[width=0.85\columnwidth,angle=0]{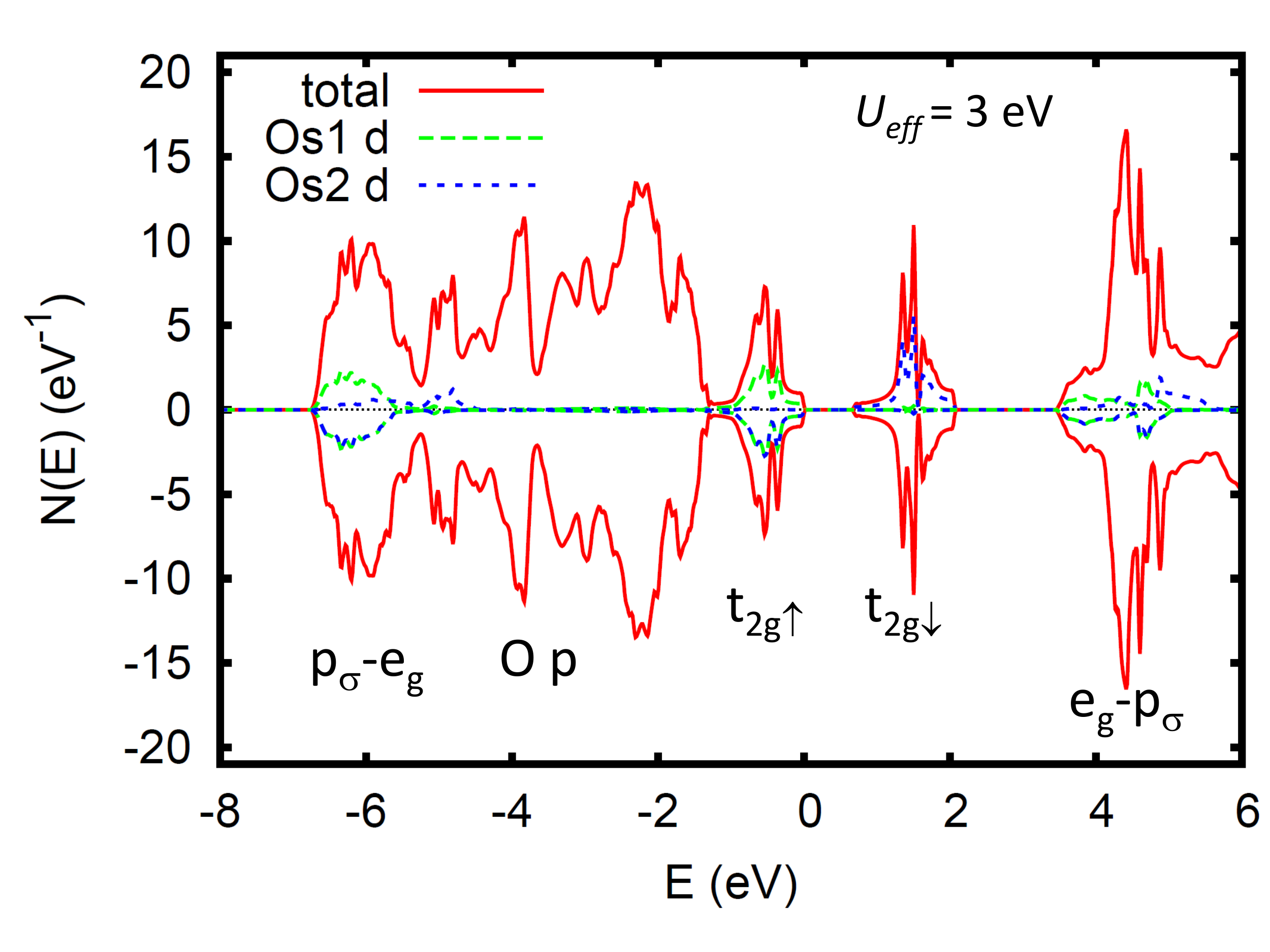}
\caption{(Color online) Calculated density of states and Os $d$ projections, from
GGA+U calculations including spin-orbit. Os1 and
Os2 denote the up and down spin Os relative to the global direction
(set along the $a$-axis).
The projections are per atom for a given spin, while the total
is per spin for the two Os atom unit cell.}
\label{dos}
\end{figure}

The density of states from the GGA+U calculation is shown in
Fig.~\ref{dos}. Even with the
inclusion of $U$ there is an exceptionally strong hybridization between
Os $5d$ and O $2p$ states evident in the projections. This hybridization provides a mechanism for relatively high
ordering temperatures despite the large separation between Os magnetic ions in the double perovskite structure. 
Specifically, it leads to a redistribution of the 
moment to include sizable moments on the O ions, with the $\sim$ 3 $\mu_B$ of
spin moment spread over OsO$_6$  that form effective magnetic clusters.
These magnetic clusters are in close proximity and are therefore strongly interacting. This is similar to the clusters found in the related double perovskite Sr$_2$YRuO$_6$ with $T_\mathrm{N}=26\,$K~\cite{mazin},
but the hybridization is stronger in the present case.

From the GGA+U calculation we find an Os moment of 1.8\,$\mu_B$, without $U$ it is only 1.4\,$\mu_B$.
The calculated orbital moment
is reasonably large, -0.11\,$\mu_B$ in both cases, but
the major reduction in the Os moment appears to result from hybridization.
This agrees with the experimental finding that the moment observed via neutron scattering is greatly reduced from the $t_{2g}$$^3$ spin-only value, but there is only a slight reduction in $\mu_\mathrm{eff}$ determined from magnetization measurements.


In this report, by combining a full experimental characterization of \SSOO{} with DFT calculations in which we include spin-orbit coupling, we have been able to establish the key properties controlling \SSOO{}. 
The Os $t_{2g}$$^3$ ground state strongly hybridizes with the surrounding oxygen ions, resulting in one of the highest magnetic transition temperatures for a double perovskite hosting a single magnetic ion. A consequence of the hybridization is the dramatically reduced ordered moment observed by neutron diffraction. Spin-orbit coupling appears to play only a minor role. This is indicated by the small reduction in the effective moment from a Curie-Weiss analysis of the susceptibility, as well the minimal effects due to spin-orbit coupling shown by DFT.

\emph{Note Added.}  After submission of this paper, an online report was published of the properties of Sr$_2$MgOsO$_6$ which contains Os$^6+$, 5$d^2$ ions~\cite{yuan_high-pressure_2015}. The magnetic susceptibility data is consistent with an AFM  transition $T_\mathrm{N}=110\,$K, and is the only reported example of a single magnetic ion double perovskite with \TN{} higher than \SSOO{}.

\section*{Acknowledgements}

Support for this research was provided by the Center for Emergent Materials an NSF Materials Research Science and Engineering Center (DMR-0820414). The research at ORNL's Spallation Neutron Source and High Flux Isotope
Reactor was supported by the Scientific User Facilities Division, Office of Basic Energy Sciences, U.S. Department of Energy (DOE). The theoretical calculations (DJS) were supported by the 
Department of Energy, Basic Energy Sciences, Materials Sciences and Engineering Division. The authors gratefully acknowledge A. Huq for providing help with
the experimental work on Powgen, and J.Q. Yan, D. Mandrus and B. Sales for useful discussions. 

\bibliographystyle{apsrev4-1}
\bibliography{Sr2ScOsO6_bib}

\end{document}